\documentclass[sigconf,screen]{acmart}
\usepackage{booktabs}
\usepackage{makecell}
\usepackage{algorithm}
\usepackage{algorithmic}
\usepackage{xurl}
\usepackage{microtype}
\usepackage{graphicx}
\usepackage{textcomp}
\usepackage{xcolor}
\usepackage{multirow}
\usepackage{xfrac}
\usepackage{tcolorbox}
\usetikzlibrary{arrows.meta}
\usepackage{balance}  

\usepackage{xcolor}
\usepackage{xspace}

\newcommand{\etal}{\hbox{\emph{et al.}}\xspace}

\newcommand{\ie}{\hbox{\emph{i.e.}}\xspace}

\PackageWarning{TODO:}{X}
\PackageWarning{TODO:}{X\%}

\usepackage{tcolorbox}
\newcommand{\Meta}{FAANGCompany\xspace}
\newcommand{\SEV}{outage\xspace}
\newcommand{\SEVs}{outages\xspace}
\newcommand{\SITEEVENT}{outage\xspace}
\newcommand{\aSEV}{an outage\xspace}
\newcommand{\ASEV}{an Outage\xspace}
\newcommand{\SEVRC}{O-RC\xspace}

\setcopyright{cc}
\setcctype{by}
\acmDOI{10.1145/3805760.3814916}
\acmYear{2026}
\copyrightyear{2026}
\acmISBN{979-8-4007-2601-9/2026/07}
\acmConference[AIware '26]{Proceedings of the 3rd ACM International Conference on AI-Powered Software}{July 6--7, 2026}{Montreal, QC, Canada}
\acmBooktitle{Proceedings of the 3rd ACM International Conference on AI-Powered Software (AIware '26), July 6--7, 2026, Montreal, QC, Canada}
\acmSubmissionID{fseaiware26main-pp016-p}
\received{2026-02-15}
\received[accepted]{2026-03-28}

\title{A Preliminary Study on Explaining Risk of Code Changes using LLM-Based Prediction Models}

\author{Yalin Liu}
\orcid{0000-0003-2199-9725}
\email{yalinliu@meta.com} 
\author{Kosay Jabre}
\orcid{0009-0003-7144-7926}
\email{kosayjabre@meta.com}
\author{Rui Abreu}
\orcid{0000-0003-3734-3157}
\email{ruiabreu@meta.com}
\author{Zachariah J. Carmichael}
\orcid{0000-0002-7603-2004}
\email{craymichael@meta.com}
\affiliation{%
  \institution{Meta Platforms, Inc.}
  \city{Menlo Park, CA}
  \country{USA}
}

\author{Vijayaraghavan Murali}
\orcid{0009-0009-1374-7334}
\email{vijaymurali@meta.com}
\author{Akshay Patel}
\orcid{0009-0000-8975-8907}
\email{akshaypatel@meta.com}
\author{Jun Ge}
\orcid{0009-0001-2519-9867}
\email{jakege@meta.com}
\author{Weiyan Sun}
\orcid{0009-0005-7627-8695}
\email{wysun@meta.com}
\author{Cong Zhang}
\orcid{0000-0001-8339-7709}
\email{conzhang@meta.com}
\affiliation{%
  \institution{Meta Platforms, Inc.}
  \city{Menlo Park, CA}
  \country{USA}
}

\author{Audris Mockus}
\orcid{0000-0002-7987-7598}
\email{audris@meta.com}
\author{David Khavari}
\orcid{0009-0002-2133-706X}
\email{dkhavari@meta.com}
\author{Nachiappan Nagappan}                            
\orcid{0000-0003-1358-4124} 
\email{nnachi@meta.com} 
\author{Peter C. Rigby}
\orcid{0000-0003-1137-4297}
\authornote{Rigby is also a professor at Concordia University in Montreal, QC, Canada.}
\email{pcr@meta.com}
\affiliation{%
  \institution{Meta Platforms, Inc.}
  \city{Menlo Park, CA}
  \country{USA}
}

\renewcommand{\shortauthors}{Liu \etal}

\begin{document}

\begin{abstract}
Predictions by machine learning (ML) and artificial intelligence (AI) models are often received skeptically unless they are paired with intelligible explanations. In the context of just-in-time defect prediction, highlighting small portions of a software change (\textit{diff})---beyond rule-based lints---where risk may be concentrated has not yet been extensively investigated.
In this work, we leverage attention weights from an LLM-based Diff Risk Score (DRS) model to highlight parts of a diff that the model focuses on when predicting risk. We aggregate token-level attention into interpretable code units (lines, hunks, and files), and present the top-$K$ units to developers as a lightweight form of guidance during code review. We evaluate our approach using expert-labeled changes that have caused real \SEVs. Results show that the highlighted snippets cover expert-labeled \SEV-causing change lines 53.85\% of the time when highlighting the top-2 hunks, while requiring developers to review 26.28\% of the changed lines on average. Because attention is produced during standard model inference, the approach is scalable for large development workflows and can be surfaced in the code review UI with low additional latency.
\end{abstract}



\maketitle

\section{Introduction}
In the (big) tech industry, software quality has a direct impact on revenue, and various tools are utilized to help maintain quality to the highest standards. A just-in-time defect prediction system (e.g., \emph{Diff Risk Score} (DRS)~\cite{abreu2024movingfasterreducingrisk,10.1145/3722216}) is employed by some companies to predict potential issues in code changes. This type of system produces a risk score indicating the likelihood of a service \SEV if a particular code change is landed. If the risk is deemed high enough, developers may be prevented from landing the change, or required to take additional steps (e.g., adding tests, requesting additional review, or performing further validation).

We have historically explored the use of logistic regression (LR) models for gating risky diffs (e.g.,~\cite{zhao2023systematic,10.1145/3722216}). These LR models are trained on curated and intuitive measures, such as the number of code review comments and the author's experience with that part of the codebase. Logistic regression is a simple causal model where each predictor has a coefficient that shows if, and by how much, it increases or decreases the risk. It has been argued that causal models are necessary for explainability~\cite{pearl2019seven,madumal2020explainable}. More complicated ML models can only tell how important a feature is to explain the variation in the response, but lack this simple mechanism provided by logistic regression~\cite{fryer2021shapley,sundararajan2020many}. Researchers have introduced entire toolkits to support traditional feature-based explainability methods~\cite{AI_Explainability360,hu2023xaitk,kokhlikyan2020captum}.

A significant limitation of the LR model is its inability to fully utilize the textual content of a given diff (such as the raw code changes themselves). In contrast, the emergence of large language models (LLMs) has opened possibilities to employ textual and other information for risk prediction. Recent work in the software engineering community, such as Diff Risk Score (DRS)~\cite{abreu2024movingfasterreducingrisk}, has shown the promise of using state-of-the-art LLMs for gating risky code changes. By risk-aligning LLMs through fine-tuning on code changes, summaries, and titles, DRS was able to gate risky diffs with significantly higher accuracy, capturing over 40\% of \SEVs by gating just the top 10\% of diffs by risk~\cite{abreu2024movingfasterreducingrisk}. It is important to note that diffs causing \SEVs are incredibly rare (approximately two in a thousand) and prediction models tend to perform poorly. Hence, inherently, the coverage is low and developers need as much explanation and justification in order for them to seriously pay attention to flagged diffs.

Despite the higher accuracy of LLMs, engineers demand to know why their diffs are considered risky~\cite{dam2018explainable}---a risk score alone is not \textit{actionable}. This need for an explanation is a major obstacle to the use of LLMs for risk prediction. The lack of explanation is particularly problematic for larger diffs, which might contain multiple large files. In addition, model errors are unavoidable: even a correct high-level warning can be ignored if it does not help an engineer narrow down what to inspect, test, or reconsider.

In this work, we present a pragmatic method to make LLM-based risk scores actionable by leveraging aggregated attention mapped to code hunks. When the model produces a risk score, we highlight which parts of the diff it attends to most strongly. We do not claim causal faithfulness between the attention signal and the model's internal reasoning; rather, we provide a low-latency highlighting signal, validated against expert-annotated outage root causes, that narrows inspection to a small number of code regions.   

Our initial results show that our method covers risky lines in the diff that actually caused faults 53.85\% of the time when highlighting the top-2 hunks. By providing pragmatic highlighting validated against expert root-cause annotations, our approach aims to address a major challenge of utilizing LLM-based risk models at scale, where the responsibility of determining what should and should not be released falls to the engineers writing and reviewing the code. Beyond increasing trust and adoption, attention-based explanations are highly scalable since attention is generated as a part of LLM inference.
Our key contribution is to develop and evaluate a pragmatic highlighting approach---aggregated attention mapped to code hunks and validated against expert-annotated outage root causes---that makes just-in-time LLM-based defect predictions actionable for professional software engineers, measurably improving over the current production baseline of no highlighting. This study is done in a large-scale industrial context at \Meta.

The remainder of this paper is organized as follows. Section~\ref{sec:background} discusses \Meta{}’s development practices and explainability challenges. Section~\ref{sec:approach} details the use of LLMs and attention weights to highlight risky code segments. Section~\ref{sec:experiment} covers the datasets, metrics, and experimental results on explanation effectiveness as well as user feedback. Section~\ref{sec:related} covers related work on explainable AI in defect prediction. Section~\ref{sec:limitations} discusses the threats to the validity of this work, and
Section~\ref{sec:conclusion} concludes with a summary and discussion of future directions.
\begin{figure*}[!t]
    \centering
    \includegraphics[width=\linewidth]{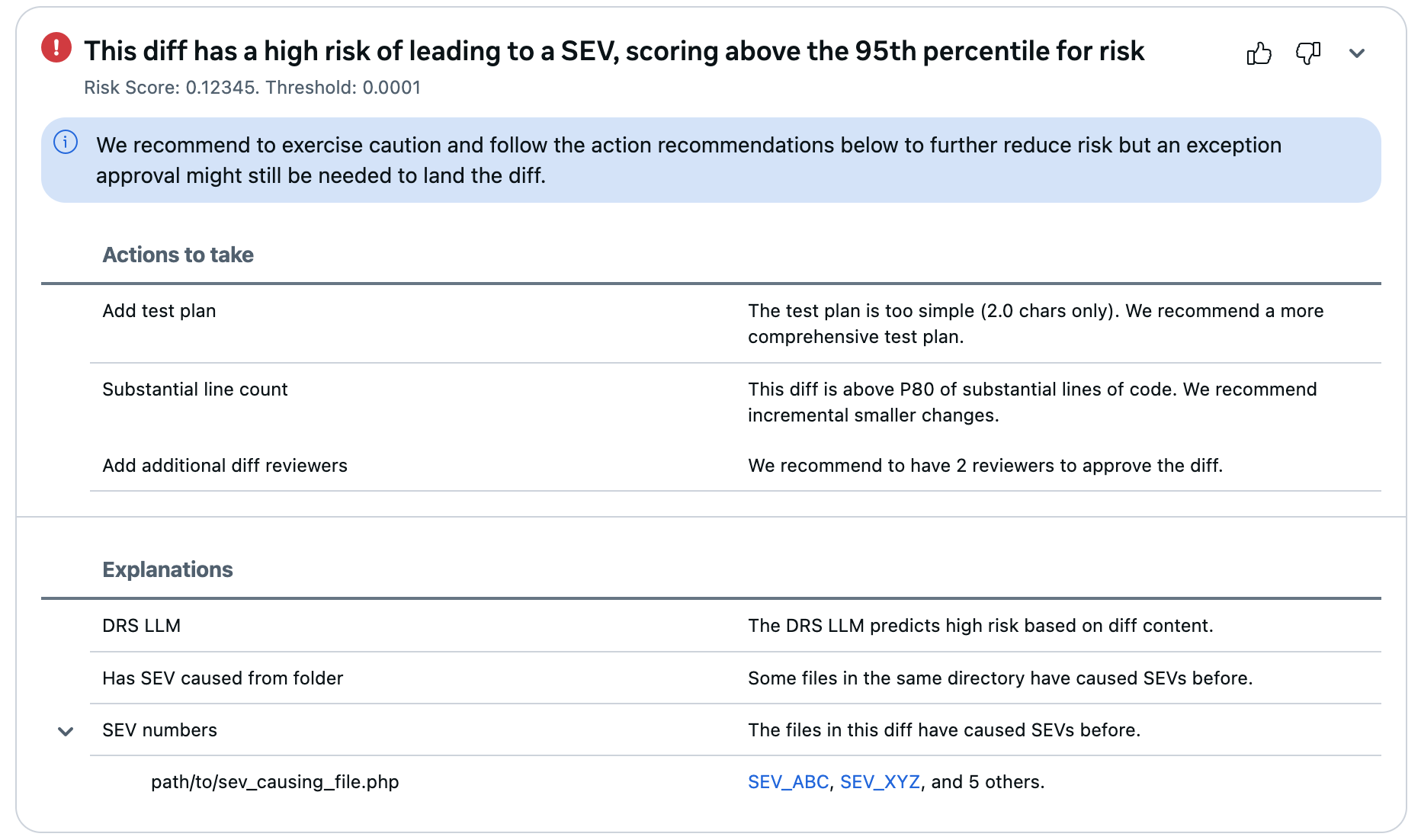}
    \caption{A redacted Code Review Tool UI showing risk and explainability of an \SEV (referred to as ``SEV'' in the image).}
    \label{fig:phab}
\end{figure*}
\section{Background}
\label{sec:background}
\subsection{Developing Software at \Meta}
\label{sec:swdev}
At \Meta, software engineers submit tens of thousands of pull requests (or ``diffs'') every day. Each diff goes through extensive automated testing, code review, canary deployment, and more to maintain a high level of code quality and reliability in our systems\footnote{What is it like to write code at \Meta? --- \Meta Tech Podcast, Episode 55, \url{https://engineering.fb.com/2023/09/05/web/what-like-ship-code-meta-tech-podcast/}}. While many diffs are routine, a small fraction can introduce severe incidents due to subtle interactions, incomplete validation, or misconfigurations.
When systems do break, we track and report these incidents as \SITEEVENT (or ``\SEVs'') as part of our incident response process. In this work, we focus on \SEVs that can be attributed to a particular landed diff, enabling root-cause analysis at the level of changed lines.

Diffs can---and do---cause \SEVs. During major revenue events, like the holiday season, we used to conduct ``code freezes'' in which we disallowed all diffs from landing in an effort to minimize \SEVs. This lowered risk at the cost of developer productivity. Recently, we have moved away from this practice toward a more dynamic risk management system that allows us to balance productivity and risk~\cite{abreu2024movingfasterreducingrisk}.

\subsection{Diff Risk Score (DRS)}
Diff Risk Score models generate a score between $0$ (no risk) and $1$ (maximum risk) for diffs before they land~\cite{abreu2024movingfasterreducingrisk,10.1145/3722216}. The score can then be used to allow the lowest risk diffs to land or block the highest risk diffs from landing (with mechanisms for exceptions in emergencies), depending on the desired productivity-risk trade-off. In practice, teams may tune thresholds and policies (e.g., block above a certain score, require additional review, or require additional testing) based on their tolerance for risk and operational constraints.
Figure~\ref{fig:phab} shows what developers see in the code review UI. It includes the risk score of the diff, reasons why diff is considered risky, and potential next actions. It also includes a feedback mechanism to let engineers provide feedback for future improvements. Our goal is to strengthen this UI so that engineers can quickly determine \emph{what} to inspect when a diff is flagged.
\subsection{Explaining Risk}
The DRS toolset is designed to be actionable and transparent. It provides feedback on why a diff received a particular risk score and suggests actions engineers can take to reduce that risk, like improving test coverage or refactoring code, as shown in Figure~\ref{fig:phab}.

A score alone is insufficient to achieve this objective. As an AI/ML based approach, DRS is often seen as a black box by software engineers. For many cases, users are informed that a diff is dangerous with little justification or context. This is especially the case when using LLM-based DRS classification as opposed to logistic regression based classification, where the latter offer some explainability through their coefficients and feature values.

The lack of good explanations presents major barriers to the adoption of DRS:
\begin{itemize}
    \item \textbf{Reduced User Productivity:} Without clear instructions, engineers struggle to unblock themselves from landing high-risk diffs, compromising developer productivity.
    \item \textbf{Reduced Trust:} Inaccurate results without rationale drive skepticism and decreased attention to warnings.
    \item \textbf{Harder Debugging and Maintenance:} Debugging DRS without rationale is like tuning a black box, making maintenance and improvement difficult.
\end{itemize}

\section{Approach}
\label{sec:approach}
To tackle the above issues, we highlight parts of the diff for engineers to inspect when a diff is flagged as high risk. In contrast to feature-based explanations that summarize \emph{why} a diff is risky in terms of metadata, we focus on \emph{where} the model attends within the diff content itself.
\subsection{Overview}
Our approach, depicted in Figure~\ref{fig:overview}, uses an LLM to evaluate the risk of a code change (diff) potentially causing severe issues. The diff is embedded into a pre-defined prompt, tokenized, and fed into the LLM, which produces a risk score based on a single logit output. During this inference, the transformer computes attention weights between the generated output token and the input tokens. We aggregate these attention weights to produce a token-level importance vector, then map token importances back to code units (lines/hunks/files). We present the top-$K$ units to developers as a compact ``review budget'' intended to reduce time spent scanning large diffs.
Because attention is computed during standard inference, highlighting adds minimal overhead and is feasible to deploy at scale in high-throughput code review workflows.
\subsection{LLM-Based Diff Risk Score Model}
\begin{figure*}[!ht]
    \centering
    \includegraphics[width=\linewidth]{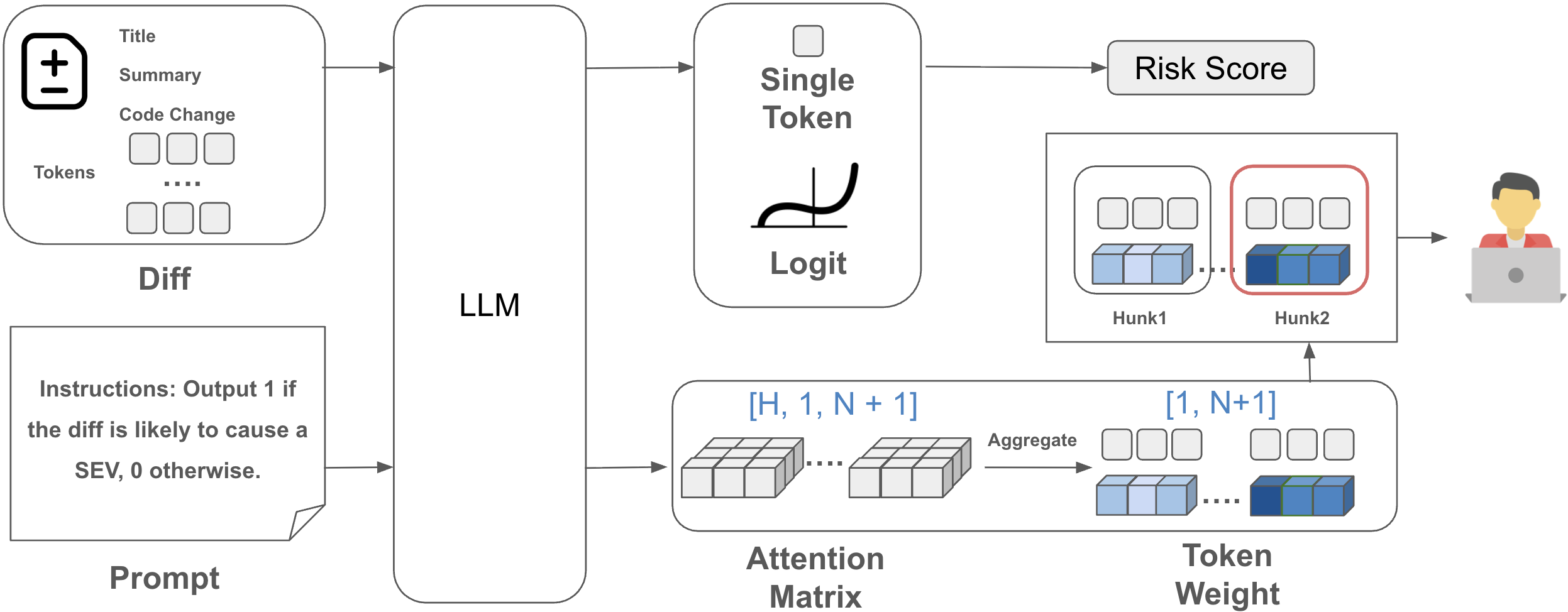}
    \caption{Overview of leveraging a LLM attention matrix to highlight code snippets worth user review. The LLM takes a pre-defined prompt to instruct the model to distinguish the risk of a diff. The LLM generates a one-token sentence and the logits of the tokens --- $0$ and $1$ --- are used to infer the diff risk score. The attention matrix is aggregated to produce a $\{1, N+1\}$ matrix representing each token's correlation towards the generated token. We further chunk the code into hunks and assign hunk importance by aggregating their tokens' attention weights.}
    \label{fig:overview}
\end{figure*}

To generate risk scores using LLMs, we apply the prompt demonstrated in Figure~\ref{fig:phab}. As shown in Figure~\ref{fig:overview}, we instruct the LLM to generate a one-token response (0 or 1) for whether or not the diff could cause \aSEV. We then take the tokens' logits as the risk likelihood. Concretely, the model produces scores for both output tokens; we transform these logits into a single risk score that can be thresholded by downstream policy (e.g., block, require additional review, or warn). The model used in this work leverages a LLM instance~\cite{touvron2023llama} which is instruction fine-tuned on historical diffs, similar to~\cite{abreu2024movingfasterreducingrisk}.
This design allows deployment with a flexible gating threshold to adapt to different base rates of \SEVs across products and infrastructure components. It also enables continuous calibration as the risk distribution changes over time. The LLM model we leverage in this work is the Llama3-70B~\cite{abreu2024movingfasterreducingrisk}.
\subsection{Explanations With Attention Weight}
\label{subsec:method_explanation}
Leveraging transformer attention weights for the purpose of explanation has been widely discussed~\cite{9897347,jain2019attention,tutek2020staying,wen2022revisiting}. In this work, we use attention as a practical proxy signal to highlight portions of the input the model attended to when producing its classification output. Our objective is to provide actionable guidance (what to inspect) while avoiding free-form natural language rationales that may hallucinate.

\paragraph{Token-level attention extraction}
Given a prompt with diff content of length $N$, the LLM generates a response of length 1. We obtain the last-layer attention tensor for the generated token with shape $\{H, 1, N+1\}$ where $H$ refers to the number of attention heads and $N +1$ is the length of input prompt plus the generated one token response. We then pool across heads by averaging to obtain a vector $w \in \mathbb{R}^{N+1}$, where $w_i$ denotes the attention weight assigned to the $i$-th input token when generating the output.

\paragraph{Mapping tokens back to code}
Token-level weights are difficult for humans to interpret directly. Therefore, we map tokens back to code units that match how developers naturally review diffs:
\begin{itemize}
    \item \textbf{Line:} individual changed lines (fine-grained but may lack context).
    \item \textbf{Hunk:} contiguous blocks of change as shown by diff tools (contextual and commonly used in review).
    \item \textbf{File:} files touched by the diff (coarse-grained; useful for triage).
\end{itemize}
We compute a unit score by aggregating token weights within the unit and then return the top-$K$ scored units. In the UI, these are highlighted to direct developer attention. Section~\ref{sec:hierarchical-grouping} goes into further detail on the grouping and aggregation approach.
\begin{algorithm}[!t]
\caption{Attention-based highlighting for diffs}
\label{alg:highlight}
\begin{algorithmic}[1]
\STATE \textbf{Input:} Prompt $P$ containing diff; chunking scheme $\mathcal{C}$; budget $K$
\STATE Run LLM on $P$ to generate a one-token output and collect last-layer attentions
\STATE Extract attention tensor $A \in \mathbb{R}^{1 \times N+1 \times H}$ for the generated token over input tokens
\STATE Pool over heads: $w_i \leftarrow \frac{1}{H}\sum_{h=1}^{H} A_{1,i,h}$ for $i \in \{1,\dots,N+1\}$
\STATE Map tokens to chunks (lines/hunks/files) using parser alignment (hierarchical token grouping)
\STATE Score each chunk as in Section~\ref{sec:aggregation}
\STATE \textbf{Return:} Top-$K$ chunks by score
\end{algorithmic}
\end{algorithm}


\subsection{Hierarchical Token Grouping}
\label{sec:hierarchical-grouping}

LLM tokenizers decompose source code into subword units
whose boundaries rarely coincide with the constructs that developers
reason about during code review.
A variable name such as \texttt{getUserById} may be split into three or
more tokens, while a hunk header or filename may be scattered across
dozens.
To produce explanations that are actionable at the granularity of
\emph{tokens}, \emph{lines}, \emph{hunks}, and \emph{files}, we
introduce a four-level hierarchy that (i)~reconstitutes subword tokens
into semantically meaningful units, (ii)~recovers the diff's structural
boundaries from tokens, and (iii)~recursively aggregates per-subword
attention scores upward through the hierarchy with level-specific
aggregation.

\subsubsection{Token Reconstitution}
\label{sec:token-reconstitution}

The first stage maps a flat sequence of subword tokens and their
associated scalar attention weights into \emph{token groupings}---units
that correspond to whitespace-delimited tokens in the original diff
text.  The DRS pipeline supports multiple backbone LLMs, so this logic is dependent on the tokenizer family.

\paragraph{SentencePiece (word-boundary merging).}
SentencePiece tokenizers mark word boundaries with a leading
\texttt{\textvisiblespace} (U+2581) prefix.  We scan the subword
sequence left-to-right, accumulating consecutive subwords into a single
token grouping until a new boundary prefix is encountered.  The prefix
character is stripped and a synthetic whitespace symbol is inserted
between groupings so that downstream formatting can reconstruct the
original text.  The special byte-token \texttt{<0x0A>}, which
SentencePiece uses for the newline character, is emitted as a dedicated
newline symbol with a null score (not considered for attribution of risk).

\paragraph{Tiktoken / BPE (pass-through).}
Tiktoken-family tokenizers encode whitespace as part of the token text
and do not use a word-boundary marker.  Each raw token is therefore
mapped one-to-one to a token grouping after splitting on literal newline
characters.  No multi-subword merging is required.

In both cases, the output is an ordered list of \emph{token grouping}
objects, each carrying the concatenated surface string and (the list
of per-subword attention scores.

\subsubsection{Hierarchical Structure Recovery}
\label{sec:structure-recovery}

Given the flat sequence of token groupings, we reconstruct the diff's
hierarchy in
two
passes.

\paragraph{Line splitting.}
Token groupings are partitioned into \emph{line groupings} by detecting
newline-delimiter tokens.


\paragraph{Hunk boundary detection.}
Within each file's code-change section, unified-diff hunk headers of the
form \verb|@@ -a,b +c,d @@| are detected via regex.  Each header
initiates a new \emph{hunk} object; subsequent lines are classified as
additions (\texttt{+}), deletions (\texttt{-}), or context based on
their leading character, and source/target line numbers are tracked
incrementally from the header's start positions.

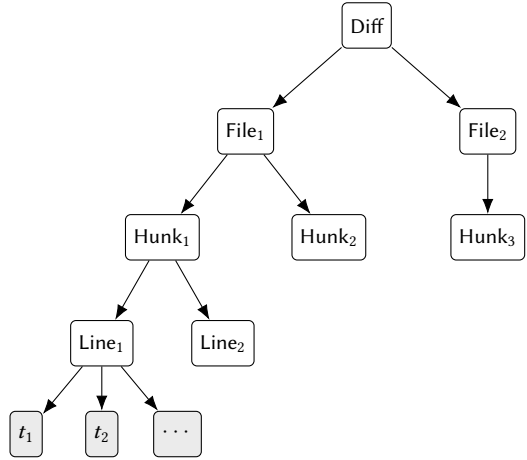
\begin{figure}[t]
\centering
\begin{tikzpicture}[
    level distance=14mm,
    sibling distance=28mm,
    every node/.style={
      draw, rounded corners=2pt, font=\small\sffamily,
      minimum height=6mm, inner sep=3pt
    },
    edge from parent/.style={draw, -{Latex[length=2mm]}},
    level 1/.style={sibling distance=32mm},
    level 2/.style={sibling distance=22mm},
    level 3/.style={sibling distance=16mm},
  ]
  \node {Diff}
    child { node {File$_1$}
      child { node {Hunk$_1$}
        child { node {Line$_1$}
          child [level distance=12mm, sibling distance=10mm] { node[fill=black!8] {$t_1$} }
          child [level distance=12mm, sibling distance=10mm] { node[fill=black!8] {$t_2$} }
          child [level distance=12mm, sibling distance=10mm] { node[fill=black!8] {$\cdots$} }
        }
        child { node {Line$_2$} }
      }
      child { node {Hunk$_2$} }
    }
    child { node {File$_2$}
      child { node {Hunk$_3$} }
    };
\end{tikzpicture}
\caption{Hierarchical grouping of diff attention scores.
Subword tokens ($t_i$) are leaves; scores are aggregated bottom-up
through lines, hunks, and files with level-specific functions.}
\label{fig:hierarchy}
\end{figure}

The result is a tree rooted at the diff, with interior nodes for files
and hunks and leaves for individual lines and tokens (depicted in Figure~\ref{fig:hierarchy}).

\subsubsection{Multi-Level Score Aggregation}
\label{sec:aggregation}

Each node in the hierarchy carries a scalar attention score.
Leaf scores are the raw per-subword weights produced by the LLM's
attention heads; interior scores are computed by a recursive,
bottom-up aggregation pass.
Because the semantics of ``importance'' differ across levels---a
subword's contribution to its parent token is qualitatively different
from a hunk's contribution to its parent file---we employ
\emph{type-dispatched} aggregation: for each node (e.g., \textit{Hunk} or \textit{Line}),
a type-specific aggregation function is selected during the recursive traversal.

Formally, let $s(v)$ denote the score assigned to node $v$ and
$\text{ch}(v)$ the ordered children of~$v$.  For a leaf token grouping
with subword scores $(w_1,\dots,w_m)$,
\[
  s(v) = \frac{1}{m}\sum_{i=1}^{m} w_i\,.
\]
For an interior node at hierarchy level $\ell$ with aggregation function
$f_\ell$,
\[
  s(v) = f_\ell\!\bigl(\{s(c) \mid c \in \text{ch}(v),\;
         s(c) \neq \texttt{null}\}\bigr).
\]
Children whose scores are null (e.g., whitespace symbols or nullified
test-file tokens) are excluded from the input multiset before
$f_\ell$ is applied.  If the resulting input is empty, $s(v)=0$.

The top-$k$ sum used at the hunk level is defined as
\[
  f_{\text{hunk}}(\mathbf{x}) =
    \sum_{i=1}^{\min(k,|\mathbf{x}|)} x_{(i)},
\]
where $x_{(1)} \ge x_{(2)} \ge \cdots$ is the sorted order of the
children's scores.

\subsubsection{Practical Considerations}
\label{sec:practical}

\paragraph{Escaped-newlines.}
Certain tokenizers split the two-character escape sequence \verb|\\n|
across token boundaries (e.g., one token for \verb|\| and a separate
token for \verb|n|).  A dedicated merge pass detects and rejoins these
fragments before line splitting, preventing spurious line breaks that
would corrupt hunk parsing.  At the hunk level, line splitting accepts
both literal and escaped newlines as delimiters to accommodate
cross-tokenizer variation.

\paragraph{Test-file score nullification.}
Code changes in test files are often mechanically correlated with the code under test and rarely represent the \emph{root cause} of a risky diff. As a pragmatic engineering choice layered on top of the raw attention signal, we nullify all token scores within files whose paths match any of a curated set of 16 regular expressions covering common test-file conventions across languages. Because null scores are excluded during aggregation, test files receive a score of zero at every level. This regex-based filtering is incomplete: remaining false highlights on auto-generated code suggest that stronger signals such as build-system metadata or codegen signatures should be incorporated in future work.


\subsection{Implementation}
Our implementation leverages the Hugging Face library\footnote{Hugging Face. (2022). Transformers: State-of-the-art natural language processing. Retrieved from \url{https://huggingface.co/docs/transformers/index}}, utilizing its unified API interface to ensure compatibility with other large language models (LLMs)~\cite{touvron2023llama}. The DRS service is implemented as a standalone Thrift API service\footnote{\url{https://thrift.apache.org/}}, where clients can send diff information, including title, summary, test, and code changes.

\paragraph{Explanation Calculation}
We embed diff information into a predefined prompt and instruct the LLM to identify \SEV risk and return a one-token response. By leveraging the logit of this token, we compute the risk score of the diff being risky (Section~\ref{subsec:method_explanation}). We also obtain the attention tensor associated with generating that token and aggregate it to compute token-level weights. These weights are then used solely for highlighting; they do not affect the risk score.

\paragraph{Tokenizer-agnostic pipeline.}
The reconstitution and structure-recovery stages abstract over
tokenizer-specific details behind a common \emph{token grouping}
interface.  Adding support for a new tokenizer requires only a new
conversion function that maps raw subwords to token groupings;
downstream aggregation and explanation rendering are unchanged.

%
\begin{figure}
    \includegraphics[width=\columnwidth]{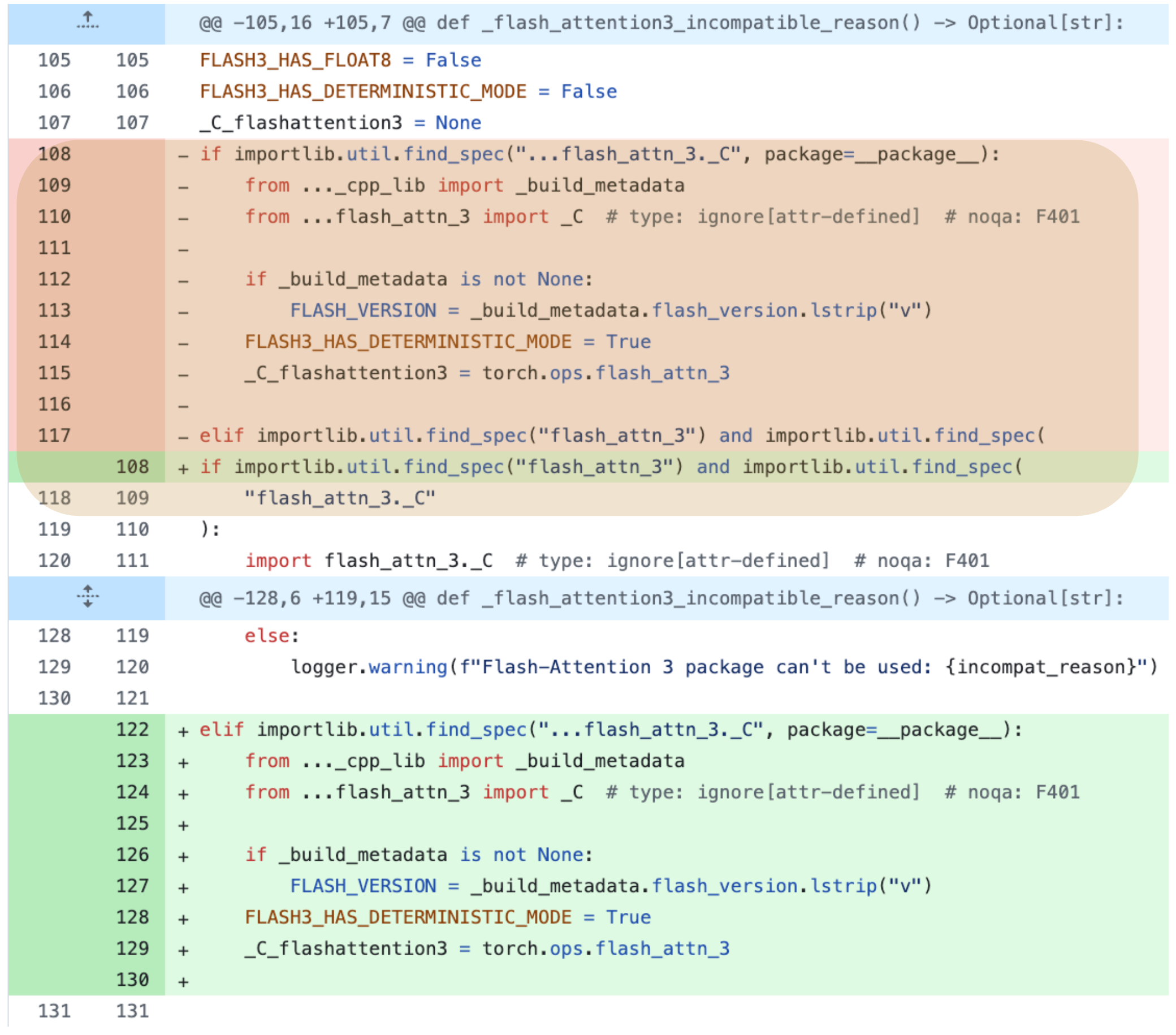}
    \caption{Illustration of hunk-level highlighting, which requests users to review 1 out of 2 hunks in a single file. The highlighted hunk contains 10 line deletion and 1 line addition.}
    \label{fig:example}
\end{figure}
\section{Experiment}
\label{sec:experiment}
\subsection{Dataset}
The method proposed in this work generates an explanation as a side product of risk score generation---it requires no additional model training. However, the DRS model is trained on data that does not include which lines caused \SEVs. As such, standard DRS model evaluation data sets are insufficient to evaluate the quality of explanations. To conduct a quantitative evaluation, we created a \ASEV Root Cause (\SEVRC) dataset that draws upon domain experts to pinpoint specific code changes that caused \SEVs.
The dataset was manually curated whereby each \SEV is tagged by four area experts. The main reviewer (typically the diff author) identifies \SEV-causing lines of the diff. The remaining three engineers perform validations. If the primary expert indicates high confidence, we include this diff in our dataset. All experts were instructed to find which lines of the diff were the most likely cause of the \SEV or leave the entry empty if it is not triggered by diff directly. The final dataset includes the file name and the code lines that caused the \SEV. For evaluation, we use our explanatory method to select the $K$ riskiest code snippets and then compute whether these snippets cover the root cause lines.
\subsection{Evaluation Setup}
\noindent
\textbf{Coverage}
This metric measures whether our method recommends results that contain the \SEV root-cause lines. Our model returns the $N$ code snippets with the highest score. We evaluate whether these snippets contain the actual \SEV-causing lines:
\[
Coverage=\frac{\text{\#diff}_{\text{root\_cause}\in\text{explanation}}}{\text{\#diffs}}
\]

\noindent          
Coverage is measured at the hunk level: a diff is counted as covered if at least one of the top-$N$ highlighted hunks contains a root-cause line. We intentionally operate at hunk granularity because it matches how diffs are reviewed in our industrial context---engineers inspect and act on hunks, not individual lines. This makes the metric coarser than a line-level measure; it captures the binary decision \emph{"does the highlighted region direct the engineer to the right hunk?"} but does not measure how precisely the highlight localizes the fault within that hunk.

\noindent
\textbf{Average Review Percent (ARP)}
Along with delivering high coverage, we seek to optimize developer time investment by asking them to review only the most important code snippets. The Average Review Percent within each diff is:
\[
ARP=avg\left(\frac{\text{\#explanation\_lines}}{\text{\#diff\_lines}}\right)
\]

We evaluate the proposed explanation method on the \SEVRC dataset to measure both attention-based highlighting accuracy and associated developer workload. As discussed in Section~\ref{subsec:method_explanation}, we chunk the diff into smaller snippets for code review. In this experiment, we explore which highlighting granularity may work best in practice. We chunk the diffs into lines, hunks and files and then return the top 10/20/30 lines for line-level highlighting, the top 1/2/3 hunks for hunk-level highlighting, and the top 1/2/3 files for file-level highlighting. Richer evaluation alternatives exist. Recall@$k$ at the line level would measure what fraction of root-cause lines fall within the top-$k$ highlighted lines. Coverage--effort curves (plotting coverage against ARP as $N$ varies) would visualize the precision/effort trade-off more completely. Area Under the Coverage-effort curve (AUCU) would provide a single aggregate measure across all operating points. We chose hunk-level coverage because it corresponds to the actionable decision in our deployment: \emph{inspect this hunk first}. The ARP metric makes the effort side of the trade-off explicit---it quantifies how much of the diff the engineer must review to achieve the reported coverage. We report results across multiple values of $N$ and granularities (line, hunk, file) so that the coverage--effort trade-off is visible, even though we do not compute a formal AUCU.
\subsection{Results / Observations}
The results of our study are presented in Table~\ref{table:result}. The line-level explanation requires a minimal amount of code review. When we return the top-10 lines to developers, they only need to review 7.6\% of the code to cover the \SEV-causing area with a coverage of 30.76\%. As the line budget increases to 30 lines, developers have more opportunity to see \SEV-causing lines in the highlighted output, at the cost of additional scanning and context reconstruction.

When we return the top-1 hunk to the user, they must review 15.37\% of the code (about 20 consecutive lines on average), which achieves 30.76\% coverage. This is worse than the top-20 line-level explanation, which requires a similar review fraction. However, top-2 hunks achieve similar results to top-30 lines on both coverage and review amount. Since hunks tend to return larger blocks of code, an incorrect prediction of the top-1 hunk may lead to more time spent in review than an incorrect prediction at the line level.

For the file-level experiment, we only include diffs where not all the files touched by these diffs cause \SEVs, \ie not all the files are part of the explanation and need to be investigated. File-level results demonstrate a trade-off: larger chunks yield higher coverage but incur substantially more review. When returning the top-1 file as the explanation, we have a coverage of 50\% with a code review percent of 41.9\% (55 lines on average). When we return the top-3 files, although we achieve 92\% coverage, users must review the majority of each diff on average (78.61\%).
Based on these observations, we find that the top-2 hunks may be the most pragmatic granularity. Although line-level explanation has higher coverage when review cost is on-par with hunk-level, it can be too fine-grained to be easily comprehended in isolation. When highlighted lines are sporadically distributed throughout a diff, users must still read surrounding lines to understand context, implicitly increasing review effort. In contrast, file-level explanations are often too coarse grained for actionable debugging or validation. Hunk-level explanation provides a balance: hunks naturally segment diffs by context and often correspond to the unit of review in real-world workflows.
By showing the top 2 hunks, we achieve 53.85\% coverage with users reviewing 26.28\% of each code change on average. We conclude that hunk 2 level provides the best balance of coverage, review lines, and review percentage.

Table~\ref{tab:agg} summarizes the aggregation choices used in production, which were selected empirically based on a held-out validation set.
We rationalize the performance of top-$k$ aggregation as it captures concentrated signal (a few high-attention lines) while
bounding the influence of very large hunks.
%

  

\begin{table}
  \centering
  \caption{Evaluation of Explanation on \SEVRC Dataset}
  \label{table:result}
  \resizebox{\columnwidth}{!}{
      \begin{tabular}{@{}llrrr@{}}
        \toprule
        \textbf{Granularity} & \textbf{Threshold} & \textbf{Precision} & \makecell[r]{\textbf{Review}\\\textbf{Lines}} & \makecell[r]{\textbf{Review}\\\textbf{Percent}} \\
        \midrule
        \multirow{3}{*}{\textbf{Line}}
        & Top 10 & 30.76\% & 10 & 7.60\% \\
        & Top 20 & 42.30\% & 20 & 15.20\% \\
        & Top 30 & 53.85\% & 30 & 22.8\% \\
        \midrule
        \multirow{3}{*}{\textbf{Hunk}}
        & Top 1 & 30.76\% & 20.22 & 15.37\% \\
        & Top 2 & 53.85\% & 34.57 & 26.28\% \\
        & Top 3 & 64.38\% & 50.19 & 38.15\% \\
        \midrule
        \multirow{3}{*}{\textbf{File}}
        & Top 1 & 50.00\% & 55.11 & 41.90\% \\
        & Top 2 & 69.23\% & 79.68 & 60.50\% \\
        & Top 3 & 92.30\% & 103.40 & 78.61\% \\
        \bottomrule
      \end{tabular}
  }
\end{table}


%
%
\begin{table}[t]
\centering
\caption{Aggregation functions by hierarchy level.}
\label{tab:agg}
\smallskip
\begin{tabular}{@{}lr@{}}
\toprule
\textbf{Level} & \textbf{Aggregation} \\
\midrule
Token        & \textsc{mean} \\[3pt]
Line         & \textsc{mean} \\[3pt]
Hunk         & \textsc{top-$k$ sum}\,(k{=}20) \\[3pt]
File         & \textsc{mean} \\
\bottomrule
\end{tabular}
\end{table}
\subsection{User Feedback}
To collect early qualitative feedback, we added our highlighting method to the code review tool and provided a user interface for positive and negative feedback with free text entry. We received 55 pieces of feedback: 7 positive and 48 negative. We treat this as early-stage debugging signal rather than a usability study---the sample is small, uncontrolled, and self-selected. Nonetheless, the negative feedback surfaced concrete failure modes that are now under active work. Among the 48 negative responses, 17 contained textual comments. These consistently pointed to three specific shortcomings: (1)~highlighting test files and auto-generated code, where our regex-based filtering (Section~\ref{subsec:method_explanation}) is incomplete; (2)~highlighting boilerplate or common patterns rather than the specific change likely to trigger an issue; and (3)~directional ambiguity---as previous studies note~\cite{liu2021exploring}, attention can support a prediction in either direction, so highlighted code may represent reasons the diff is viewed as \emph{less} risky, not more. These are known limitations.
We are actively exploring stronger filtering signals (build metadata, codegen signatures) for~(1), and methods to identify attention signal polarity for~(3).
\section{Related Work}
\label{sec:related}
The advent of exceedingly accurate AI models and their use in science and engineering has raised fundamental questions around what understanding means and how to attain it. Notably, AI models are accurate oracles, but do not, without additional technology, help humans understand their predictive rationale. One pragmatic solution from the philosophical standpoint~\cite{de2005contextual} is that two conditions are satisfied:
\begin{enumerate}
\item A phenomenon P can be understood if a theory T of P exists that is intelligible (and meets the usual logical, methodological, and empirical requirements).
\item A scientific theory T is intelligible for scientists (in context C) if they can recognise qualitatively characteristic consequences of T without performing exact calculations.
\end{enumerate}
In case of DRS predictions, the key question is to what extent the highlighted risky parts of the diff are ``intelligible'' to an engineer making decisions about the change.

According to Krenn \etal~\cite{krenn2022scientific}, an AI system can contribute to new scientific understanding as a ``computational microscope'' with the ability to acquire information not yet attainable through experimental means; as a ``resource of inspiration'' or an artificial muse, expanding the scope of human imagination and creativity; as an ``agent of understanding'', replacing the human in generalizing observations and transferring new scientific concepts to different phenomena, and conveying these insights to human scientists. In the case of DRS, the LLM without highlighting serves as a microscope showing which diffs may cause outages, but it becomes a more useful ``agent of understanding'' when it can also indicate where to inspect.

Explanation is key for understanding~\cite{de2020understanding}, and engineers and scientists are not satisfied without it. AI methods have a long history of dealing with explainability. From early rule-based expert systems through to decision trees, explainability typically took the form of revealing underlying rules and branch weights. Newer methods like random forest obscured the rules, but one could still estimate the importance of each feature used in prediction. Deep learning upended previous approaches to explainability. According to Xu \etal~\cite{xu2019explainable}, the primary approach with deep learning models is post-hoc explanation: a result is inferred then an explanation is generated. Post-hoc explanation tries to (a) provide analytic statements; (b) show saliency maps via feature importance---in our case, to highlight a subset of a change; (c) give explanations by example.
In the context of defect prediction, previous work includes fitting simple models to explain more complex models~\cite{esteves2020understanding}. That might look like predicting risk with an LLM, but explaining those predictions with a logistic regression model.

A systematic review of the literature on explainable AI in software engineering~\cite{mohammadkhani2023systematic} shows that defect prediction is the most common target, but that explainability is mostly investigated for traditional rather than deep learning methods~\cite{yu2024formal}. Hence, our contribution is both relevant and timely. The survey finds that visualization and explanations in natural language are particularly uncommon in the field of explainable AI for software engineering. 

Note, however, that the focus of this paper is to show that our approach is useful at scale at \Meta{}, rather than a systematic comparison with other approaches.
\section{Limitations}
\label{sec:limitations}
\paragraph{Generalizability}
Generalizing from empirical studies in software engineering is difficult because of a large number of potentially relevant context variables. The study was performed at \Meta, so the results might be different elsewhere. However, the software systems under study involve millions of lines of code and thousands of developers some of whom are collocated while others collaborate across multiple locations across the world. The systems also span a range of domains from social network products and virtual and augmented reality to software engineering infrastructure.
While the desire to prevent outages is common among all (big tech) companies, the specific DRS tool we use may be unique to this study at present, but it could be of great value more broadly. Our (preliminary) evaluation represents only a snapshot in time; the distribution may shift over time and for other domains or types of risk.
\paragraph{Construct Validity}
We have used two measures: the existence of overlap with triggering entities and the percent of code highlighted. Even experts who annotated our diffs may not be 100\% accurate as to what exact lines caused the outage; we plan to conduct future studies with more users to determine if the approach is effective in practice.
We assume that highlighting helps developers identify and fix problems, but we have not yet measured downstream outcomes such as reduced time-to-resolution or reduced post-landing incident rates. We plan to conduct follow-up studies to understand how developers interpret highlighted snippets and how that interpretation affects decision-making.
\paragraph{Internal Validity}
We make an assumption that the attention accurately reflects areas of the code that need inspection, but there may be better ways to indicate and explain the risk. We intentionally avoided some approaches we deemed too risky such as asking the model to provide natural language explanation that may produce hallucinations or factually incorrect claims. Additionally, attention weights may be influenced by prompt structure or formatting, and not solely by semantic properties of the change.
\section{Conclusions \& Future Work}
\label{sec:conclusion}
This paper presented how LLMs are used at \Meta to generate a score of how likely a diff will lead to an outage---a system we call Diff Risk Score (DRS)~\cite{abreu2024movingfasterreducingrisk}. In this paper, by leveraging attention weights, we proposed a pragmatic highlighting approach that uses aggregated attention, mapped to code hunks, to direct engineers toward the parts of the change most associated with the predicted level of risk.. We experimented at multiple levels: line, hunk, and file. To balance the coverage against the number of lines that an engineer needs to review, our initial results show that displaying the top-2 hunks to engineers is sufficient. We found a coverage of $53.85$\% while highlighting an average of $34$ lines accounting for $26$\% of the total lines in a diff.

While our work is preliminary, our key contribution is the development and validation of a pragmatic highlighting signal based on aggregated attention and expert-annotated outage root causes. This signal makes just-in-time LLM-based defect predictions actionable for professional engineers, measurably improving over the current production baseline of no highlighting. This approach may increase engineer trust in the model and help them mitigate risks associated with their changes.. This has the potential to improve the reliability and trustworthiness of DRS models, enabling their broader deployment in a variety of industry settings. Furthermore, our attention-based highlighting is highly scalable and efficient for real-world, large-scale software development workflows (such as the one at \Meta).

Future work will focus on refining this approach, exploring its applications in other domains, and evaluating its impact on both software quality and developer productivity. Promising directions include incorporating additional signals (e.g., file history or test coverage) to reduce false highlights, and developing methods that distinguish attention associated with a high-risk prediction from attention associated with the opposite.

\balance
\bibliographystyle{ACM-Reference-Format}
\bibliography{references}

\end{document}